\RequirePackage{fix-cm}
\documentclass[smallextended]{svjour3}       % onecolumn (second format)
\smartqed  % flush right qed marks, e.g. at end of proof
\usepackage{graphicx}
\usepackage{amsmath}
\usepackage{amssymb}
\usepackage{multirow}
\usepackage{amsfonts, mathabx}
\usepackage{enumitem}
\usepackage{graphicx,kantlipsum,setspace}
\usepackage{caption}
\usepackage{xr-hyper} 
\usepackage{hyperref} 
\usepackage{natbib}
\usepackage[ruled, vlined]{algorithm2e}

\newcommand{\E}{\mathbb{E}}
\newcommand{\pr}{\mathbb{P}}

\newcommand{\R}{\mathbb{R}}
\newcommand{\Z}{\mathbb{Z}}

\newcommand{\mc}{\mathcal}

\newcommand{\vep}{\varepsilon}

\newcommand{\heta}{\wh{\eta}}

\newcommand{\bx}{\mathbf{x}}

\newcommand{\cG}{\mathcal{G}}

\newcommand{\beas}{\begin{eqnarray*}}
\newcommand{\eeas}{\end{eqnarray*}}

\def\wh{\widehat}
\def\wt{\widetilde}
\begin{document}

\title{Ensemble Binary Segmentation for irregularly spaced data with change-points\thanks{DISCLAIMER: The views and opinions
rendered in this article reflect the author's personal views about the subject and do not necessarily represent the views of
any affiliation.}
}


\author{Karolos K. Korkas
}


\institute{University of Greenwich,  School of Computing and Mathematical Sciences.\\
              \email{kkorkas@yahoo.co.uk}   
}

\date{Received: date / Accepted: date}

\maketitle

\begin{abstract}
We propose a new technique for
consistent estimation of the number and locations of the
change-points in the structure of an irregularly spaced time series. The core of the segmentation procedure is the Ensemble Binary Segmentation method (EBS), a technique in which  a
large number of multiple change-point detection tasks using the Binary Segmentation (BS) method are applied on sub-samples of the
data of differing lengths, and then the results are combined to create an overall answer.
We do not restrict the total number of change-points
a time series can have, therefore, our proposed method works well when the spacings between change-points are short. Our main change-point detection
statistic is the time-varying Autoregressive Conditional Duration model on which we apply a transformation process in order to decorrelate it. To examine the performance of EBS we  provide a
simulation study for various types of scenarios. A proof of consistency is
also provided. Our methodology is implemented in the R package
\verb+eNchange+, available to download from CRAN.
\end{abstract}

\section{Introduction}
\setlength{\parskip}{0.0mm}
Irregularly spaced time series, i.e. data recorded as and when they emerge, have attracted  significant attention due to the development and increasing automation in information technology. This type of data, aka high-frequency data, has not only penetrated financial markets, whereby trade transactions, as a result of continuous electronic trading, have generated a vast amount of datasets. Examples of irregularly spaced time series can be found in many industrial and scientific domains such as in natural disasters where floods, earthquakes or volcanic eruptions typically occur at irregular time intervals; in clinical trials where a patient's state of health is typically observed at different points of time; or in data collecting sensors (smart applications of Internet of Things) which are activated only when an event happens to, e.g., preserve battery energy.

Modelling high-frequency data successfully requires that the dynamics in the time series data are captured efficiently and standard time series methods are not appropriate for these data. To address this, a class of models, the autoregressive conditional duration models (henceforth, ACD), was originally introduced by \cite{engle1998ACD}. In particular, the authors model the conditional mean as an autoregressive process which has a multiplicative relationship with positive valued error terms. These specifications resemble the multiplicative structure of the GARCH model used for modelling an asset's return volatility, but thanks to the positive valued error terms ACDs are more suitable for modelling the dynamics of continuous positive valued random variables, such as durations (the time it takes between two trades occurring in a trading book), trading volume (the size of orders), or market depth (the available liquidity at any point of time). ACD models have become popular with many variations owing partially to the popularity of the GARCH model and its numerous extensions. 
%Without being exhaustive, we refer the reader to Weibull ACD, gamma ACD, and generalized gamma ACD; to models with more flexible dynamic specification such as the logarithmic ACD model of \cite{bauwens2000logarithmic}, the Box-Cox ACD model of \cite{dufour2000time}; and to non-parametric or semi-parametric versions found in \cite{cosma2006nonparametric} or \cite{saart2015semiparametric}. 
An alternative to ACD's autoregressive specification is to parameterize the intensity function which assumes a self-exciting process. This class of stochastic process was originated by \cite{hawkes1971} and, hence, these stochastic processes are referred to as Hawkes processes. They serve as epidemic models: the occurrence of a number of events (e.g. seismic or buy/sell) will increase the probability of further events. Their use in modelling financial high-frequency data has picked up with many applications and variations.

In practice, however, time series entail changes in their dependence structure and the above mentioned models are more appropriate for stationary time series. It will be a crude approximation to adopt them in modeling the non-stationary processes without at least a minimal adaptation in a model's assumptions. There is also high risk in prediction and forecasting in doing so, as noted by \cite{mercurio2004} for financial data. Dealing with high-frequency financial data is not less risky. A representative example is the U-shape in a stock's trading activity typically observed during a day. We cannot expect a market's behaviour to be like for like at every point of time. Information flows almost continuously and, therefore, the intra-day dynamics vary significantly and cannot be ignored.

The simplest type of deviation from stationarity is, arguably, piecewise stationarity. This implies that the parameters of a stochastic process remain constant (hence, the process is stationary) for a certain period of time. In this paper, we focus on this type and we aim to identify the stationary segments by detecting the locations and number of change-points. If, of course, we knew the segments a-priori we would fit a stationary model to each of the segments and proceed with the prediction or forecasting tasks, but this knowledge is typically absent.

Detecting change-points has attracted significant attention. One approach to solve this problem is to formulate it through an optimization task i.e. minimising a multivariate cost function (or criterion) and adding a penalty when the number of change-points is unknown (see \citep{yao1988} or \citep{davisetal1995} among others). However, this approach typically comes with a high computational cost. To reduce the computational complexity \cite{killick2012optimal} proposed PELT that has linear computational time under the assumption that the number of change-points increases linearly with sample size.

%Depending on the model a user can adopt certain cost functions: the least squares for change in the mean of a series \citep{yao1989} or \citep{lavielle2007adaptive}, the Minimum Description Length criterion (MDL) for non-stationary time series \citep{davisetal1995}, the Gaussian log-likelihood function for changes in the volatility \citep{lavielle2007adaptive} or the covariance structure of a multivariate time series \citep{lavielle2006detection}.

%Change-point detection methods that adopt a multivariate cost function often come with a high computational cost. Dynamic programming (\cite{bellman1966} and \cite{kay1998}), Segment Neighbourhood \citep{auger1989algorithms} or Optimal Partitioning \citep{jackson2005}, are used in solving change-point problems, but their complexity is at least $\mathcal{O}(T^2)$ where $T$ is the sample size.

An alternative approach to the change-point estimation problem is to minimize a series of univariate cost functions in a `greedy' manner, i.e. detect a change-point and then progressively move to identify more. A popular representative of this category is the Binary Segmentation method (BS) and the reasons for its popularity are its low computational complexity and easiness of implementation: after identifying a single change-point (through the use of a certain statistic such as the CUSUM) the detection of further change-points continues to the left and to the right of the initial change-point until no further changes are detected.

The BS algorithm has been adopted to solve various types of problems with \cite{inclan1994} to be, perhaps, the first to use it to detect breaks in the variance of a sequence of independent observations. 
%\cite{kim2000} extend \cite{inclan1994} method to a GARCH(1,1) model and \cite{lee2001} extend the same method to linear processes. \cite{piotr2011} use the BS method to test for multiple change-points in an ARCH process. \cite{cho2012} apply the binary segmentation method on the wavelet periodograms in order to identify change-points in the second-order structure of a nonstationary process. Using the wavelet periodogram, \cite{killick2013wavelet} propose a likelihood ratio test under the null and alternative hypotheses, but assume that the number of change-points are bounded from above. The BS method is also used for multivariate (possibly high-dimensional) time series segmentation in \cite{cho2013} for detecting change-points in the second-order structure and in \cite{cho2018} for detecting change-points in high-dimensional multivariate GARCH processes. 
For irregularly spaced time series, at least in the context of ACD or Hawkes processes, no literature exists that proposes a BS method (or an alternative change-point detection method) to detect change-points in a piecewise stationary ACD or Hawkes process. We note that \cite{roueff2016locally} introduce the time-varying Hawkes process which is locally stationary, not necessarily piecewise stationary and, hence, they do not deal with change-point detection.

The BS algorithm, under specific model specifications, may be inefficient in detecting the change-points. \cite{fryzwild} illustrates that with a simple signal+noise model having only three change-points close to each other and in the middle of the series. At the beginning of the BS algorithm the CUSUM statistic does not clearly point to a change-point, hence, it does not move to search for more. This behaviour should not come as a surprise: BS is a greedy algorithm searching for a single change-point at each iteration and failing to do so results in an unnecessary early stop. The Wild Binary Segmentation algorithm (WBS \cite{fryzwild}) aims to solve the above mentioned limitation using a `certain random localization mechanism'. We discuss WBS later on, but we note here that this randomized algorithm was extended to univariate time series \citep{korkas2017} and high-dimensional panel data \citep{wang2018high}. Other variants of the random localization mechanism include the second version of  WBS (WBS2) of \cite{fryzlewiczWBS2} and the Narrowest-Over-Threshold (NOT) method of \cite{baranowski2016}, while \cite{anastasiou2019} propose Isolate-Detect (ID) which involves a fixed (rather than random) localization mechanism making it of an order of magnitude faster than the methods mentioned above. A faster implementation of WBS is SeedBS of \cite{kovacs2020seeded} that relies on a deterministic construction of the localization mechanism.

In this work, we capitalize on BS's popularity and propose a new randomised version of it which we term Ensemble Binary Segmentation (EBS). In particular, we draw a number $M$ of random segments from a given univariate time series and apply the BS algorithm in each of these segments. The estimated change-points are then collected from over the $M$ BS applications. Due to this simple mechanism the ways to combine the estimated change-points are numerous, for instance, the final set of change-points comprise of those that appear frequently or appear more frequently relative to other estimated change-points. An extra feature is that the estimated change-points can be ranked based on their relative frequency which is useful when post-processing change-points.

The reader might question why WBS is not preferred for detecting change-points in a piecewise stationary ACD  or Hawkes process and a new method is proposed instead. First, EBS exhibits better actual performance both in terms of computational speed and controlling the total number of change-points. The way that WBS works does not allow an efficient post-processing of the estimated change-points resulting in sometimes significant oversegmentation (in the case of a few change-points) or undersegmentation (in the case of frequent change-points). The source of the problem is the spurious detection of change-points which is a consequence of the distributional features of the ACD multiplicative form. Further, EBS, in practice, is fast because it does not search for a single change-point in every iteration, but rather it can locate all the change-points with even a single random draw with a small, albeit non-zero, probability. Second, contrary to WBS, the method we propose here can be extended to include other segmentation algorithms. In other words, when drawing a number of $M$ of random segments we do not have to apply BS in each of these, but any other method can be potentially used. Therefore, studying EBS is a crucial starting point for other ensemble-type of change-point detection techniques.

The action of aggregating many method outputs to randomized versions of the data is not new. In fact, it has been studied extensively in the statistics and machine learning literature. Random forest is a popular representative that enjoys good predictive performance. Briefly, a random forest works as follows: grow multiple regression (decision) trees to randomized parts of the training  dataset, and then average the trees. Our proposed method works in a similar manner and provides a fresh solution to the change-point problem.

The paper is structured as follows: In Section \ref{sec:tvACD_intro} we
present the time-varying ACD model (tvACD), the core of our
detection algorithm, and in Section \ref{sec:Stage_A} its transformation. The EBS algorithm is presented in Section \ref{sec:Stage_B}, including its
theoretical consistency in detecting the number and locations of
change-points. The same section covers the post-processing of the estimated change-points and conducts simulation studies to obtain universal algorithm parameters that almost minimize the need for tuning by the user. Further, in  Section \ref{sec:Simulat} we conduct a simulation study to
examine the actual performance of our proposed algorithm vis-\`a-vis the standard BS algorithm. In Section \ref{sec:Apps} we apply our
method to financial high-frequency transaction data. Proofs of the results are in the supplementary material. Our methodology is implemented in the
R package \verb+eNchange+ (\citealp{korkas_ebs_r_package}), available to download from CRAN.

\section{Time-varying ACD model}\label{sec:tvACD_intro}

Let $\tau_t$ be the transaction time where $0= \tau_0 <...<\tau_{T-1}$ and $x_t=\tau_t - \tau_{t-1}$ be the duration between trades/events. We consider the following time-varying ACD model
\begin{eqnarray}
x_t / \psi_t &=& \varepsilon_t \; \sim \cG(\theta_1)\label{eq:acd:one}
\\
\psi_t &=& \omega(t) + \sum_{j=1}^p \alpha_{j}(t)x_{t-j} + \sum_{k=1}^q \beta_{k}(t)\psi_{t-k}.
\label{eq:acd:two}
\end{eqnarray}
where $\psi_{t} = \E [x_t | x_{t-1},...,x_{t-p},\psi_{t-1},...,\psi_{t-q}; \Theta(t)]$ is the conditional mean duration of the $t$-th event with parameter vector $\Theta(t)$ and $\cG(.)$ is a general distribution over $(0,+\infty)$ with mean equal to 1 and parameter vector $\theta_1$. The vector of parameters involved in the conditional
mean $\psi_{t}$ is
$$\Theta(t) = (\omega(t), \alpha_j(t), \beta_j(t)) \in \mathbb{R}^{1+p+q}.$$
Then, $\Theta(t)$ is piecewise constant in $t$ with $N$
change-points $\eta_i$ ($\eta_0 =0,
\eta_{N+1} = T$) i.e. at any $\eta_i$ we have that
$\Theta(\eta_i) \neq \Theta(\eta_i+1)$ for all $i=1,\cdots, N$. The number of change-points  and their locations are allowed to increase with $T$, hence, formally, we have that $N=N(T)$ and $\eta_i=\eta_i(T)$ for $\forall i=1,\cdots, N$. To save notation and be consistent with the existing literature on change-point detection, henceforth, we use the shorthand notation $N$ and $\eta_i$.

The number of change-points $N$ and their locations are assumed to
be unknown and we aim to estimate them.  The parameter values in
$\Theta(t)$ are also unknown, but these can be estimated after the
change-points have been detected and stationary segments are
identified.

We further assume the following conditions on (\ref{eq:acd:one}) and (\ref{eq:acd:two})
\begin{enumerate}[label=(A\arabic*), start=1]
\setlength\itemsep{0em}
\item\label{eq:a1} For some $\epsilon_1 > 0$, $\xi_1 < \infty$ and all
$T$, we have
\begin{eqnarray*}
\inf_{1 \leq t \leq T}\omega(t) > \epsilon_1 \mbox{ and } 
\sup_{1 \leq t \leq T}\omega(t) \le \xi_1 < \infty.
\end{eqnarray*}

\item\label{eq:a2} For some $\epsilon_2 \in (0, 1)$ and all $T$, we have
\begin{eqnarray*}
\sup_{1 \leq t \leq T} \{\sum_{j=1}^p\alpha_{j}(t)+\sum_{k=1}^q\beta_{k}(t)\} \le 1-\epsilon_2.
\end{eqnarray*}
\end{enumerate}

Assumptions \ref{eq:a1}--\ref{eq:a2} 
guarantee that between any two consecutive change-points, 
$x_{t}$ admits a well-defined solution a.s. and is weakly stationary (see e.g., Theorem 4.35 of \cite{douc2014}).

For a non-stationary stochastic process $x_t$, its strong-mixing rate is defined as a sequence of coefficients $\alpha(\kappa)$ such that
\begin{eqnarray}
\label{eq:alpha_mix}
\alpha(\kappa) = \sup_{t\in\Z} \sup_{\substack{G \in \sigma(X_{t+\kappa}, X_{t+\kappa+1},\ldots), 
\\ H \in \sigma(X_t, X_{t-1}, \ldots)}} 
|\pr(G \cap H) -\pr(G)\pr(H)|. 
\end{eqnarray}

In \cite{piotr2011} the mixing rate of univariate, time-varying ARCH processes was investigated, and in \cite{cho2018} the mixing rate of any  pair of time-varying GARCH processes was established. 
A tvACD process is also strong mixing at a geometric rate,
under the following Lipschitz-type condition on 
the density $f_{\vep_{t}}$ of $\vep_{t}$. 

\begin{enumerate}[label=(A\arabic*), start=3]
\setlength\itemsep{0em}
\item\label{eq:a3} The  distribution of $\vep_{t}$ satisfies the following: 
for any $a>0$, there exists fixed $K > 0$ independent of $a$ such that 
\beas 
\int | f_{\vep_{t}}(u) - f_{\vep_{t}}(u[1+a]) | du \le Ka.
\eeas 
\end{enumerate}
\vspace{5pt} 

\cite{piotr2011} show that \ref{eq:a3} is satisfied when certain conditions about a density function $f$ hold i.e. the first derivative is bounded; after some finite point $m$ the derivative $f'$ declines monotonically to zero; and $\int_0^{\infty} |zf'|dz < \infty$. It is straightforward to show that $f_{\vep_{t}}$ satisfies these conditions, hence, \ref{eq:a3} holds when $\varepsilon_t \sim \exp(\theta_1)$. Many well-known distributions satisfy this condition (see \cite{piotr2011}), but extension of our method to these will require a transformation of the duration process that accounts for the various distribution parameters. This is a technically challenging task, and it is left for future research. In this work, we assume that distribution $\cG(\theta_1)$ in (\ref{eq:acd:one}) is $\exp(1)$.

%Then, time-varying bivariate GARCH processes $\br_{ii', t}$, $i < i'$, is strong mixing as below 
%\begin{prop}
%\label{prop:one} Under \ref{eq:a1}--\ref{eq:a3}, there exists some $\alpha \in
%(0, 1)$ such that
%\begin{eqnarray*}
%\sup_{1 \le i < i' \le N}\sup_{\substack{G \in \sigma(\br_{i, i', u}: \, u \ge t+k),
%\\ H \in \sigma(\br_{i, i', u}: \, u \le t)}} | \pr(G \cap H) - \pr(G)\pr(H) | \le M\alpha^k,
%\end{eqnarray*}
%where $M$ is a finite constant independent of $t$ and $k$.
%\end{prop}
%See Appendix A.2 in the supplementary document for the proof.

A different approach to modeling the arrival times is through the Hawkes process: a self-exciting process where the arrival of an event (or events) causes the conditional intensity function $\lambda({\tau})$ to increase. In this work, we introduce the  time-varying Hawkes process (which we term \emph{tvHawkes}). In particular, we consider a counting process on $\mathbb{R}^{+}$  $\{\mathcal{N}_{\tau}\}_{\mathcal{N}>0}$ with associated history $\{\mathcal{F}_{\tau}\}_{{\tau}\geq 0}$ whose conditional intensity is defined as
\begin{equation}\label{eq:tvHawkes}
\lambda({\tau}) = \lambda_0({\tau}) +\sum_{{\tau}_t < u} \sum_{j=1}^p \alpha_j({\tau})
e^{-\beta_j({\tau}) ({\tau}-{\tau}_t)}, \ \mbox{for} \ {\tau} = 1, \ldots,T^h.
\end{equation}
where $\lambda_0 > 0$ is the initial intensity at time ${\tau}=0$ and $\alpha_j({\tau})$, $\beta_j({\tau})$ are positive parameters that are piecewise constant in $\tau$ with $N$ change-points $\eta_i$. We refer the reader to the supplementary material for more details.

To prove consistency of EBS in detecting the number of change-points and their locations when the underlying model is a tvHawkes process we would need assumptions around non-negativity of $(\lambda_0({\tau}), \alpha_j({\tau}), \beta_j({\tau}))$ and stationarity in between any two consecutive change-points. However, to the best of our knowledge, no mixing rates have been established for a time-varying Hawkes process and, therefore, it is not trivial to estimate an upper bound for the cumulative error term. In this work section, we mainly focus  on tvACD while we use tvHawkes as a benchmark (and an alternative) to tvACD.

%\section{Two-stage change-point detection methodology}\label{sec:main_algorithm}

\section{Transformation of the duration process}\label{sec:Stage_A}
In the first stage of our proposed methodology we form the process $U_t$ using a function $g_0: \R^{1+p+q} \to \R$ that takes $\bx_t^{t-p} = (x_t,x_{t-1},\cdots,x_{t-p})^T$ and $\boldsymbol{\psi}_{t-1}^{t-q} = (\psi_{t-1},\cdots,\psi_{t-q})^T$ as inputs. This function $g_0$ is required to be bounded and Lipschitz continuous.

\begin{enumerate}[label=(A\arabic*), start=4]
\setlength\itemsep{0em}
\item\label{eq:a4} The function $g_0: \R^{1+p+q} \to \R$ satisfies 
$|g_0| \le \bar{g} < \infty$ and is Lipschitz continuous,
 i.e., $|g_0(z_0, \ldots, z_{p+q}) - g(z_0', \ldots, z_{p+q}')| \le C_g\sum_{k=0}^{p+q}|z_k-z_k'|$.
\end{enumerate}

Empirical residuals are widely
adopted for detecting changes in the parameters of a stochastic
model and in this work it will form the basic statistic for our
detection algorithm in the context of point processes.

For the tvACD model, following \cite{piotr2013} we select $g_0$ such that
\begin{eqnarray}
U_t &=& g_0(\bx_t^{t-p},\boldsymbol{\psi}_{t-1}^{t-q}) =
\frac{x_t}{\widecheck{\psi}_t}, \quad \mbox{where}
\label{eq:gone}
\\
\widecheck{\psi}_t &=& C_0 + \sum_{j=1}^p C_j x_{t-j}
+ \sum_{k=1}^q C_{p+k} \psi_{t-k}+\epsilon x_t
\nonumber
\end{eqnarray}
where the last term $\epsilon x_t$ is added to ensure the boundness of $U_t$. This transformation decorrelates the original tvACD process and lightens its tails. Therefore, it serves as our main change-point detection statistic by observing that when any parameter $\Theta(t)$ undergoes a change at some point $t$, so does $\E (U_t)$. In this work we favour $\log(U_t +\epsilon)$ where $\epsilon$ as in (\ref{eq:gone}) to reduce the rightward skew observed in the distribution of $U_t$. The log-transformation does not impact the mixing properties of the process and, as a result, our consistency result for $U_t$ still holds for $\log(U_t +\epsilon)$ (see, e.g., Theorem 14.1 in \citep{davidson1994stochastic}).

It is important to note that $x_t$ or any `diagonal' transformation of $x_t$ such as $\log(x_t^2)$ should not be used instead of $U_t$ because both $x_t$ and the logarithmic process  $\log(x_t^2)$ are highly autorrelated. This will distort the change-point detection most likely by producing a false picture of the locations of the change-points. For a discussion on transformations we refer to the original work of \cite{piotr2013} and for the tvHawkes transformation to the supplementary material.

We prepare the ground for a consistency result for our proposed method. Let $\{bx_t^i\}_{t=1}^T$ denote a stationary ACD process with parameters $\Theta(\eta_i +1)$, and the innovations coinciding with $\varepsilon_t$ over the associated segment $[\eta_i+1,\eta_{i+1}]$. For each $i=1,\cdots,N$ we also define $\wt{U}^{i}_{t} = g_0(\bx_{t}^{i, t-p},\boldsymbol\psi_{t-1}^{i, t-q})$ and denoting the index of the change-point strictly to the left and nearest to $t$ by $v(t)=\max\{0\leq i \leq N: \eta_i <t\}$ with which $\{x_t^i\}_{t=1}^T$ and $\wt{U}^{i}_{t}$ are defined.

\begin{proposition}
\label{prop:two}
Suppose that \ref{eq:a1}--\ref{eq:a4} hold, and let $y_{t}=U_{t}$. Then, we have the following decomposition
\begin{eqnarray}
\label{eq:panel} 
y_{t} = f_{t} + z_{t}, \quad 1 \le t \le T.
\end{eqnarray}
\begin{itemize}
\item[(i)] $f_{t}$ are piecewise constant as
$f_{t} = \wt{g}_{t}$ 
where
$\wt{g}_{t} = \E\{(\wt{U}^{v(t)}_{t})\}$.
All change-points in $f_{t}$ belong to $\mc B = \{\eta_1, \ldots, \eta_N\}$.
\item[(ii)] $z_{t}$ satisfies
\beas
\max_{1 \le s < e \le T}
\frac{1}{\sqrt{e-s+1}}\left\vert \sum_{t=s}^e z_{t} \right\vert
 = O_p(\sqrt{\log\,T}).
\eeas
\end{itemize}
\end{proposition}

Proof of Proposition \ref{prop:two} can be found in the supplementary material. Unlike $\E(U_{t})$, 
we have $\wt{g}_{t}$ which is
{\em exactly} constant between any two adjacent change-points
without any boundary effects.
By its construction, $z_{t} = U_{t} - \E(\wt{U}^{v(t)}_{t})$ 
does not satisfy $\E(z_{t}) = 0$, but thanks to the mixing properties of $U_{t}$ as in (\ref{eq:alpha_mix}), 
its scaled partial sums can be appropriately bounded.
In the next section, we introduce the multiple change-point detection algorithm.

\section{The segmentation algorithm}\label{sec:Stage_B}

\subsection{The Binary Segmentation algorithm}
We first present the Binary Segmentation algorithm within the framework
of the ACD model. In the next section, we explain how this algorithm can be enhanced by ensembling detected change-point from multiple applications of the BS algorithm on smaller (and random) segments.

We consider the CUSUM-type statistic which has been widely adopted for
change-point detection in both univariate and multivariate data
\begin{eqnarray}
\label{eq:single_cusum}
\mathbb{Y}_{s,e}^b =
\sqrt{\frac{(b-s+1)(e-b)}{e-s+1}}\left(\frac{1}{b-s+1}\sum_{t=s}^b
y_{t} - \frac{1}{e-b}\sum_{t=b+1}^e y_{ t} \right)
\end{eqnarray}
for $s \le b < e$. A large value of $|\mathbb{Y}_{s,e}^b|$ typically indicates the presence of a change-point in the level of $y_{t}$ in the vicinity of $t = b$. In particular, if $\mathbb{Y}_{s, e}>\pi_{T}$ where
\begin{eqnarray}
\mathbb{Y}_{s, e} = \max_{s \le b < e} |\mathbb{Y}_{s,e}^b|, \label{eq:test:stat} 
\end{eqnarray}
and $\pi_{T}$ is a threshold (the choice of which is discussed in Section  \ref{sec:thresh}), then the location of the change-point is estimated as
\beas
\heta = \arg\max_{s \le b < e}  \mathbb{Y}_{s,e}^b.
\eeas

The BS algorithm  starts by initialising with $s = 1$, $e = T$ and $\wh{\mc B} = \emptyset$ and proceeds by recursively applying the CUSUM statistic (\ref{eq:single_cusum}) on $[s,\heta]$ and $[\heta+1,e]$. The BS
algorithm stops in each current interval when no further
change-points are detected, the obtained CUSUM values fall below
threshold $\pi_{T}$.

%\begin{algorithm}[htbp]
%\caption{{\tt BinSeg} (Binary Segmentation algorithm)}
%\label{alg:bs}
%\DontPrintSemicolon
%\SetAlgoLined
%\KwIn{$\{y_{t}\}$, $\pi_{T}$, $s$, $e$, $\wh{\mc B}$}
%\BlankLine
%{\bf Step 1:} compute $\mathbb{Y}_{s,e}^b$ for $s \le b < e$;
%
%{\bf Step 2:} set 
%\beas
%\mc \mathbb{Y}_{s,e}^b \leftarrow \max_{s \le b < e} |\mathbb{Y}_{s,e}^b| \text{ and }
%\wh\eta \leftarrow \arg\max_{s \le b < e} \mathbb{Y}_{s,e}^b
%\eeas
%
%{\bf Step 3:} \If{$\mathbb{Y}_{s, e} > \pi_{T}$}{
%$\wh{\mc B} \leftarrow \wh{\mc B} \cup \{\wh\eta\}$\;
%
%$\wh{\mc B} \leftarrow$ {\tt BinSeg}($\{y_{t}\}$, $\pi_{T}$, $s$, $\wh\eta$, $\wh{\mc B}$)\;
%
%$\wh{\mc B} \leftarrow$ {\tt BinSeg}($\{y_{t}\}$, $\pi_{T}$, $\wh\eta+1$, $e$, $\wh{\mc B}$)\;
%}
%\BlankLine
%\KwOut{$\wh{\mc B}$}
%\end{algorithm}

\subsection{The Ensemble Binary Segmentation algorithm}

Our recommended enhancement of the BS algorithm, which we argue to be a significant improvement, is based on the fact that the BS method can possibly fit the wrong
model when multiple change-points are present as it searches the
whole series. The application of the CUSUM statistic
can result in spurious change-point detection when
e.g. the true change-points occur close to each other. Especially,
\cite{olshen2004} notice that BS method can fail to detect a small change in the middle of a large segment which is illustrated in \cite{fryzwild}. 

%re-write the next paragraph to avoid repetition
To solve this, \cite{fryzwild} proposes a randomised version of the  binary segmentation method (termed
Wild Binary Segmentation -- WBS) where the search for
change-points proceeds by calculating the CUSUM statistic in
smaller segments whose length is random. By doing so, WBS aims to draw favourable intervals with probability tending to one with the sample size containing at most a single change-point. To improve upon WBS mainly in reducing complexity, \cite{fryzlewiczWBS2} suggests a new recursive algorithm, termed Wild Binary Segmentation 2 (WBS2), for producing  a ‘complete’ solution path i.e. a sequence of estimated nested models containing $0,\cdots,T-1$ change-points. To achieve this he computes all possible CUSUM statistics in any given interval when it is not too long; otherwise it computes the CUSUM statistics in random segments as in WBS. The final selection of the change-points is done through the `Steepest Drop to Low Levels' procedure that is not penalty-based, and only uses thresholding as a certain discrete secondary check (WBS2.SDLL). Other solutions to the BS issue described above are that of \cite{baranowski2016} with NOT and \cite{anastasiou2019} with ID.

We provide reasons for choosing EBS over WBS or WBS2. First, both WBS and WBS2 are tailored around BS, and it is not always straightforward to apply to other segmentation methods. On the other hand, EBS can be easily extended as it collects estimated change-points from applying BS to many random sub-samples. In addition, it is typical to oversegment a series under the multiplicative setting which in turn requires further processing of the estimated change-points. How well this post-processing will perform depends on the detection ability of WBS itself both in localisation accuracy and the total number of change-points identified.  EBS, as we explain shortly after, is naturally equipped with a `second chance' mechanism of selecting the change-points and is less likely to oversegment, which gives it an advantage before entering the post-processing stage.
Because oversegmentation does not arise in the signal+iid Gaussian noise setting of \citep{fryzwild}, as it is entirely due to the distributional
features of the multiplicative setting where the input sequence is a rightward skewed and autocorrelated random sequence, the complete solution path produced by WBS2.SDLL  will not be much of use in post-processing either. The reason is that in order to rank the change-points estimated from WBS2.SDLL we would need to rank the CUSUM statistics, which in turn can take large values when the starting and ending points are close to a change-point. A solution could be to adjust for a segment's (random) length or average the CUSUM statistics for each change-point (again after adjusting for the random lengths of the segments), however, we favour EBS for being independent of the location statistic when ranking the change-points.

We propose a randomised algorithm, but instead of drawing random intervals, calculate the CUSUM statistic in each of these and proceed in a ``to-the-left-and-to-the-right'' manner, we run $M$ multiple BS on random segments of the underlying univariate series. We then collect all the obtained change-points and calculate the frequency of the occurrences by simply counting the number of times a certain change-point appears over $M$ draws of the BS algorithm. This results in a better performance compared with BS as it is more likely to draw favourable intervals which can contain a single change-point (as in the WBS methodology) or more than one (as in the BS methodology). By doing so we aim to balance the benefits of both worlds. The action of ensembling (hence, borrowing from the machine learning literature we term our methodology Ensemble BS or EBS)  allow us, through a slightly randomised mechanism, to classify a change-point as important when it appears more often than other change-points. 

In the last stage, we have the options to either i. inspect the histogram of the estimated change-points; or ii.  to keep only the change-points that appear more than $\lceil \pi_thr \cdot M \rceil$ times where $\pi_thr \in [0,1]$; or iii. to apply (ii) and then post-process the detected change-points by removing change-points that are ranked lower and/or are `close' to high ranked change-points.

As an illustration, we simulated a tvACD model of the following form
\begin{eqnarray}\label{model:main_ill}
x_t&=& (\omega(t) + 0.1x_{t-1} + 0.7\psi_{t-1}) \cdot \varepsilon_t \; \sim \exp(1)
\end{eqnarray}
with sample size $T=3000$ and $\omega(t) \in \{1/16,1/4,1/16,1/4,1/16\}$ at $.475T,.485T,.495T,\; \text{and} \; .505T$. 

\begin{figure}[htbp]
\centering
\includegraphics[scale=.27]{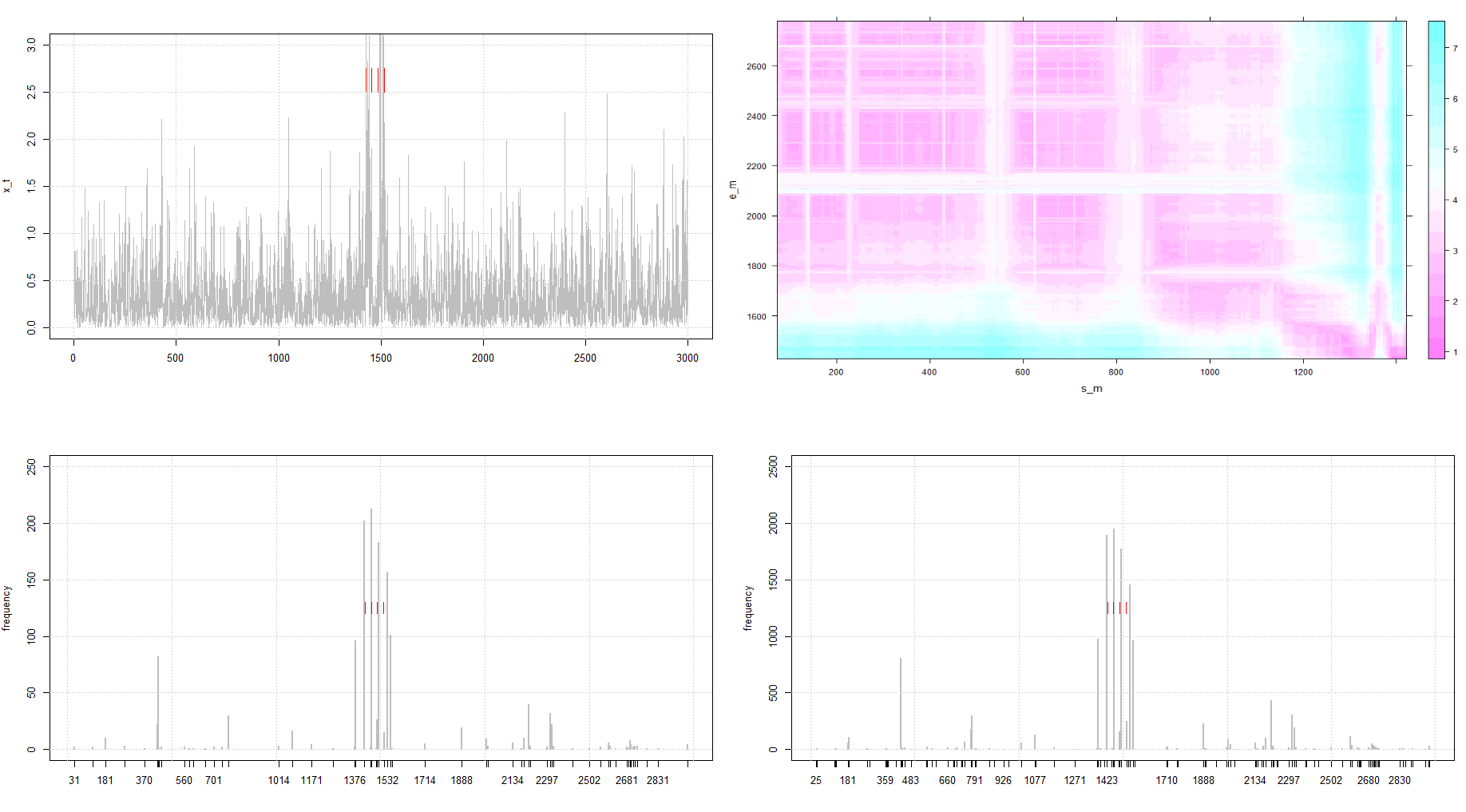}
\caption{\footnotesize{A simulated tvACD process from the model (\ref{model:main_ill}) (top left). A heatmap of the maximum of the absolute CUSUM statistics calculated on the transformed series $U_t$ with varying start ($s_m$ on x-axis) and end ($e_m$ on y-axis) points (top right). The empirical histogram of the estimated change-points from $M=1000$ draws (bottom left). The empirical histogram of the estimated change-points from $M=10000$ draws (bottom right). The four vertical bars in red are the real locations of the change-points.}}
\label{plot:main_ill}
\end{figure}

We chose this setup because change-points in the middle of a long time series is challenging for BS to detect. The CUSUM statistic (in absolute value) will fail to exceed the threshold and the BS algorithm will stop. On the contrary, when taking random intervals $(s_m,e_m)$ and then calculating the CUSUM statistic it was more likely for it to obtain a high CUSUM statistic in absoluve values (hence, more likely to exceed the threshold), see the right top plot in Figure \ref{plot:main_ill}. What is important to observe is that only $s_m$ has to be  close to the first change-point in order for the CUSUM statistic to surpass the threshold, while end point $e_m$ can be much further to the right (observe the blue area from 1200 onwards in the x-axis). This observation holds symmetrically i.e. if $e_m$ is close to the last (fourth) change-point then $s_m$ can freely take any value in the interval $[1, \lceil 0.475T \rceil)$ (observe the blue area below 1600 in the y-axis). In those both favourable scenarios (blue areas) the BS algorithm will proceed to identify the rest change-points consistently.

It is interesting to note that the number of draws $M$ did not alter the change-point performance. From the bottom two plots in Figure \ref{plot:main_ill} one can see that the shapes of the empirical distributions look identical even though in the first case $M=1000$, while in the second $M$ was 10-fold. The reason for that lies in the proof of the consistency result (see the supplementary material), but it can be easily seen by inspecting the model (\ref{model:main_ill}): a favourable draw can occur when \emph{only} the starting point $s$ is in a local neighbourhood to the first change-point from the left, and applying a single BS routine will result in to identify all adjacent change-points. Contrary to WBS, whereby it would seek to localise again after the identification of a change-point, EBS would likely identify all the change-points from the first (favourable) draw.

We now describe the EBS algorithm in more detail. First, denote by $\mathcal{D}_m=\{\wh\eta_1^{(m)}, \wh\eta_2^{(m)},\ldots\}$ for $m=1,\ldots,M$ the set of all the change-points detected by the BS algorithm on a random interval $[s_m,e_m]	\subseteq [1,T]$. Let $\mathcal{E} = \bigcup _{m=1}^M \mathcal{D}_m $, the ensemble collection of all the detected change-points from the $M$ applications of BS, and $|\mathcal{E}|$ its cardinality. Evaluate each  change-point $\wh\eta_i$ by counting the number of times (frequency) it appears in the ensemble $\mathcal{E}$, i.e.

\begin{eqnarray}\label{eq:MV}
V_{\wh\eta_i}^{(M)} = \sum_{m =1}^{M} \mathbb{I}(\wh\eta_i \in  \mathcal{D}_m) \; \text{for} \; i=1,\cdots,\wh N
\end{eqnarray}
where $\mathbb{I}(.)$ is the indicator function. In the machine learning literature, formula (\ref{eq:MV}) is referred as majority voting. We can also obtain the relative frequency of a change-point $\eta_i$ using $f_{\wh\eta_i}=V_{\wh\eta_i}/|\mathcal{E}|$ to create a relative importance plot (histogram). To make a decision about how many change-points are finally selected (and, hence, omit falsely detected change-points), one way is to select from the ensembles that $\wh\eta_i$ such that 
\begin{eqnarray} \label{eq:major_vote_thresh}
V_{\wh\eta_i} \geq \pi_{thr} \cdot M, \; \pi_{thr} \in [0,1].
\end{eqnarray}

Ensembling the change-points from a random interval sampling scheme is obviously not the only way to go. From a theoretical (and practical) point of view, EBS can be based on a deterministic grid of intervals as, for example, in \citep{kovacs2020seeded} whose segmentation method leads to a near-linear time approach. A referee pointed out that a random scheme is likely not appropriate because of the overlapping random intervals. However, it is only with a random interval sampling scheme that we can get a meaningful `aggregation' result, whereby the frequency of a detected change-point across $M$ runs can be seen as measure of its importance.

We emphasize that other types of thresholding can be utilized, for instance, a change-point is selected when it is ranked above the average, i.e.
\begin{eqnarray*}
V_{\wh\eta_i} > \frac{1}{\hat{N}}\sum_{k'=1}^{\hat{N}}V_{\wh\eta_{k'}} \; \forall k'=1,\ldots, \wh{N}.
\end{eqnarray*}

The relative importance plot also provides a type of `scree plot', whereby an elbow (kink) indicates what the right number of selected change-points is. This type of scree plot is common in, for example, Principal Component Analysis and the selection of the number of principal components.

In this work, we opt for the majority voting type of selection and we point out that the  selection based on ranking  lays ground for further research. We explain why by looking at the problem of change-point detection through the lens of variable selection (see, e.g., \citep{harchaoui2010}). We identified two views in randomised versions of variable selection. \cite{meinshausen2010} argue that a variable is declared predominant if a slightly randomized routine selects it `more frequently" than other variables (i.e. majority votes), and a threshold parameter exists that controls exactly what we mean by ‘more frequently’. However, \cite{xin2012} argue that variable ranking is more practical than variable selection as such, and the ability to rank variables  based on how often they are selected is a major advantage of the ensemble approach making it more attractive than other selection algorithms (see also \citep{zhu2015} for an interesting review of majority voting in variable selection).

Finally, one of the referees has noted that the use of (\ref{eq:MV}) to rank the detected change-points according to their importance might not be appropriate. However, we stress that a change-point that ranks high within $\mathcal{E}$ should be preferred over other change-points, but only (and optionally) for post-processing the change-points.  

\begin{algorithm}[thbp]
\caption{{\tt EnBinSeg} (Ensemble Binary Segmentation algorithm)}
\label{alg:ebs}
\DontPrintSemicolon
\SetAlgoLined
\KwIn{$\{y_{t}\}$, $\pi_{T}$, $\pi_{thr}$, $\wh{\mathcal{E}},\wh{\mathcal{B}}, M$}
\BlankLine
{\bf Step 1:} \For{$1 \ldots M$} { 

$[s_m,e_m] \leftarrow U(1,T)$

$\wh{\mc B}^{(m)} \leftarrow$ {\tt BinSeg}($\{y_{t \in {s_m:e_m}}\}$, $\pi_{T}$, $s_m,e_m,$  $\wh{\mathcal{E}}$)\;

$\wh{\mathcal{E}} \leftarrow \wh{\mathcal{E}} \cup \wh{\mc B}^{(m)}$\;
}

{\bf Step 2:} \For{$i=1 \ldots \wh{N}$} { 

\uIf{$\wh{\eta}_i \in \wh{\mc B}^{(m)}$} {
$v_{m,\wh{\eta}_i} \leftarrow 1$}
\Else{
$v_{m,\wh{\eta}_i} \leftarrow 0$}
}
{\bf Step 3:} \For{$i=1 \ldots \wh{N}$} {
$V_{\wh{\eta}_i}=\sum_{m=1}^M v_{m,\wh{\eta}_i}$
\BlankLine
\uIf{$V_{\wh{\eta}_i} > \pi_{thr} \cdot M$} {
$v_{m,\wh{\eta}_i} \leftarrow 1$\;
$\wh{\mc B} \leftarrow  \wh{\eta_i}$}
}
\BlankLine
{\bf Step 4 (Optional for post-processing):} 
Rank $\wh{\eta_i}$ by $V_{\wh{\eta}_i}$ for $i=1,\ldots, \wh{N}$
\BlankLine
\KwOut{$\wh{\mc B}$}
\end{algorithm}

%
%
%We remark that, for a sufficiently large $M$, EBS and WBS are similar because EBS will select a significant number of favourable draws $[s_m,e_m]$ each containing a single change-point. Applying the BS method on each of these intervals is equivalent to applying the CUSUM statistic once. However, EBS provides more flexibility around model selection (i.e. number and locations of change-points) and can be extended to other segmentation methods.

\subsection{Theoretical properties of EBS}\label{sec:TheoremEBS}

In this section we present the consistency theorem for the EBS
algorithm for the total number $N$ and locations of the
change-points $0 < \eta_1 < ... < \eta_N < T-1 $ with $\eta_0=0$
and $\eta_{N+1}=T$. To achieve this, we impose the following conditions in addition to \ref{eq:a1} to \ref{eq:a4}	mainly to control the detectability of each $\eta_b$.

\begin{enumerate}[label=(B\arabic*), start=1]
\setlength\itemsep{0em}
\item\label{eq:b1}  The number of change-points $N$ in (\ref{eq:acd:one})-(\ref{eq:acd:two}) is unknown and allowed to
increase with $T$ and only the minimum distance between the
change-points can restrict the maximum number of $N$.

\item\label{eq:b2}  There exists a fixed constant $f^* > 0$ such that
$ \max_{1 \le t \le T} |f_{t}| \le f^*$.

\item\label{eq:b3}  The distance between any two adjacent change-points
satisfies $\min_{r=1,...,N+1}|\eta_r - \eta_{r-1}| \geq \delta_T$,
where $\delta_T \ge C \log  T$ for a large enough $C$.

\item\label{eq:b4} The magnitudes of the change-points satisfy $\inf_{1\leq i
\leq N}|f_{\eta_{i}+1} - f_{\eta_i}| \geq
f_{\star}$ where $f_{\star}>0$.

%\item\label{eq:b5} $\Delta_T \asymp \delta_T$ where $\Delta_T>0$ is the maximum size of the interval drawn.

\end{enumerate}

\textbf{Theorem 1} \emph{Suppose that Assumptions (A1)-(A4) and (B1)-(B4) hold.
With the number of change-points as $N$ and the
locations of those change-points as $\eta_1,...,\eta_N$, let
$\hat{N}$ and $\hat{\eta}_1,...,\hat{\eta}_N$ be the number and
locations of the change-points (in ascending order) estimated by
the Ensemble Binary Segmentation algorithm. There exist constants
$C_1$ and $C_2$ such that if $C_1 \sqrt{\log T} \leq \pi_T \leq C_2
\sqrt{\delta_T}$, then  $P(\mathcal{Z}_T) \geq  1 - \pi_{z} -(1-\delta_T T^{-1}/9)^M$, where}
$$\mathcal{Z}_T = \{\hat{N} = N; \;\; \max_{i=1,...,N}|\hat{\eta}_i-\eta_i| \leq C \log T \}$$
\emph{for certain $C$ and $\pi_{z} =  \bar{C}_1  T^2\exp(-\bar{C}_2(c'\sqrt{\log\, T} -\bar{C}_3(p+q)N\log(T)^{-1/2})^2)$ for certain $c', \bar{C}_1, \bar{C}_2$ and $\bar{C}_3$. The guaranteed speed of convergence of
$P(\mathcal{Z}_T)$ to $1$ is no faster than $(1-\delta_T T^{-1}/9)^M$ where $M$ is
the number of random draws.}

\medskip

The rate of convergence for the estimated change-points obtained
for the BS method by \cite{piotr2013} is
$\mathcal{O}(\sqrt{T} \log ^ {\bar{\alpha}}T)$ where $ \bar{\alpha}>0$ when
$\delta_T$ is of order $T$. In the EBS setting,
the rate is square logarithmic when $\delta_T$ is of order $\log
 T$, which represents an improvement. 

We now elaborate on the minimum number $M$ of random draws
required to ensure that the bound on the speed of convergence of
$P(\mathcal{Z}_T)$ to 1 in Theorem 1 is suitably small. Suppose
that we wish to ensure that
\[
(1 - \delta_T  T^{-1} / 9)^M \le T^{-1}.
\]

This is equivalent to
\[
M  \ge \frac{ 9  T} {\delta_T}  \log(T)
\]
by noting that $\log(1-y) \approx -y$ around $y=0$.

Let us consider the ``easiest'' case, i.e. $\delta_T \sim T$. This results in a
logarithmic number of draws, which leads to particularly low
computational complexity, and it also has the same complexity with the WBS case. When $\delta_T \sim \sqrt{\log(T)}$, then the required $M$ increases almost linearly, but it has less computational complexity than WBS, which also explains why EBS is generally faster than WBS. 

Finally, we discuss $\pi_{thr}$ which appears in (\ref{eq:major_vote_thresh}), and it acts as a decision rule when aggregating estimated change-points across $M$ applications of the BS algorithm. In practice, EBS tended to return spurious change-points due to the distributional features of the multiplicative setting which require us to control the partial sums of $z_t$ in (ii) of Proposition 2 achieved through an appropriately chosen $\pi_{thr}$. Theoretically, it is not trivial to control the falsely detected change-points, as the distribution of the partial sums of $z_t$ is not easy to obtain, and we have to rely on simulations to identify an appropriate choice of $\pi_{thr}$. Our practical recommendations for the choice of $\pi_{thr}$ along with the choice of $M$ are discussed in Section \ref{sec:thresh}.

\subsection{Post-processing}\label{EBS.post.process}
In practice, the real number of change-points is not known to us and to reduce the risk of over-segmentation we propose a post-processing method. The need for post-processing the estimated change-points from a detection routine is common within the context of multiplicative models. We refer the reader to \cite{inclan1994}, \cite{cho2012} and \cite{korkas2017}. In these works, the post-processing method compares every detected change-point from the main detection method against the adjacent ones (re)using the CUSUM statistic. Even though there are variations in implementing a post-processing between them, their common drawback is the lack of information about the importance of the detected  change-points.

Our proposal in filtering estimated change-points is simple: starting with the highest ranked $\wh{\eta}_i^{(1)}$ based on its $f_{\wh{\eta}_i}$, we remove from set $\wh{\mc B}$ $\wh{\eta}_i^{(2)}$ if it is within a distance $\Delta_T$ and examine whether $\wh{\eta}_i^{(3)}$ is within this distance. If $\wh{\eta}_i^{(2)}$ is not within distance, we keep this change-point and repeat the process until all change-points are separated by at least $\Delta_T$. Because $\Delta_T$ cannot be smaller than the minimum permitted distance $\delta_T$ - as in Assumption \ref{eq:b3} - between any two adjacent change-points, hence, $\Delta_T=\mathcal{O}(\delta_T)$. In practice, we take $\Delta_T=\lceil 0.005 \cdot T \rceil$.

\subsection{Choice of parameters for transformation}
The right choice of the transformation function $g_0$ will influence the empirical performance of our methodology and its power in particular. This choice boils down to determine the coefficients $C_{j}, \, j = 0, \ldots, p+q$. 

The BASTA--res algorithm of \cite{piotr2013} 
performs change-point detection in the univariate ARCH process, as already seen similar to the ACD process,
by analysing the transformation of the input time series
obtained similarly to $U_{t}^2$. 
They recommend the use of `dampened' versions of the GARCH parameter estimates which we also adopt here for the ACD process. This leads to the choice of $C_{0} = \wh\omega$,
$C_{j} = \wh\alpha_{j}/F, \, 1 \le j \le p$ and $C_{p+k} = \wh\beta_{k}/F, \, 1 \le k \le q$, 
with $F \ge 1$. 

Empirically, the motivation behind the introduction of $F$ is as follows. 
For $x_{t}$ with time-varying parameters, we often observe that 
$\alpha_{j}$ and $\beta_{k}$ are over-estimated in the sense that
$\sum_{j=1}^p\wh\alpha_{j} + \sum_{k=1}^q\wh\beta_{k}$ is close to one, especially when dealing with real data. There has been evidence in the literature that change-points may cause persistence estimation in volatility models (e.g. \cite{francq2001}). \cite{mikosch2004}, among others, show that the estimated persistence close to unity in GARCH models is likely spurious and confounded by neglected change-points. We observed the same phenomenon when trying to fit an ACD process to simulated and real data.
Using the raw estimates in place of $C_{j}$'s in \eqref{eq:gone}, therefore,  is not the best approach. \cite{cho2018} propose to choose the dampening factor $F$ as
\begin{equation}\label{eq:F_dampen}
F = \max\left[1, \frac{\min(0.99, \sum_{j=1}^p\wh\alpha_{j} +
\sum_{k=1}^q\wh\beta_{k})} {\max\big\{0.01, 1 -
(\sum_{j=1}^p\wh\alpha_{j} + \sum_{k=1}^q\wh\beta_{
k})\big\}}\right].
\end{equation}

By construction, $F$ is bounded as $F \in [1, 99]$ and 
approximately brings $\wh\omega$ and $\sum_{j=1}^p\wh\alpha_{j} + \sum_{k=1}^q\wh\beta_{k}$ to the same scale. The selection of the order $p,q$ can be done through the means of an information criterion; however, taking into consideration the possibility that the estimated persistence will be close to unity, we conducted a simulation experiment to establish the right choice of order. For brevity, the details of the experiment are provided in the supplementary material.

The results indicate that the transformation involving the empirical estimates from an ACD(0,1) model vis-a-vis ACD(1,1) had a better performance. The dampening factor $F$ did not have as a strong impact and generally the performance remained unchanged for increasing $F$. In a separate experiment not shown here the dynamic selection of $F$ worked better and is, hence, recommended.  

%In this section, we do not discuss the case of a tvACD process with $p,q>1$ and we postpone it until the simulation study. There, we see that the choice of fitting a tvACD(0,1) model is adequate in detecting the change-points for a tvACD of any order.

\subsection{Choice of threshold and parameters}\label{sec:thresh}
In this section we present choices of the parameters involved in
the main Theorem $1$. In particular, we have that the
threshold $\pi_T$ includes the constant $C_1$. To approximate the distribution of the CUSUM statistic in the absence of change-points one approach is to use a parametric resampling procedure that was described in \cite{cho2018} and which provides good results. A similar approach has widely been adopted in the change-point literature including \cite{kokoszka2002} who test for the presence of a change-point in the parameters of univariate GARCH models.
%\begin{algorithm}[htbp]
%\caption{Resampling algorithm for threshold selection (tvACD)}
%\label{alg:boot}
%\DontPrintSemicolon
%\SetAlgoLined
%\KwIn{$\{x_{t}\}$, $\wh \psi_{t}$, $R$, $g_0$, $\alpha$}
%\BlankLine
%{\bf Step 1:} compute the empirical residuals $\wh\vep_{t} \leftarrow \wh{\psi}_{t}^{-1/2} x_{t}$\;
%
%{\bf Step 2:} \For{$\ell = 1, \ldots, R$}{
%{\bf Step 2.1:}
%generate bootstrap samples $\{\bm\vep^\ell_t\}_{t = 1}^T$
%of $\{\wh{\bm\vep}_t\}_{t = 1}^T$\;
%
%{\bf Step 2.2:} simulate the ACD process
%\beas
%x^\ell_{t} = (\psi^\ell_{t})^{1/2}\vep^\ell_{t}, \quad
%\mbox{where} \quad \psi^\ell_{t} = \wh{\omega} +
%\sum_{j=1}^p\wh{\alpha}_{j}(r^\ell_{t-j})^2 + \sum_{k=1}^q \wh{\beta}_{k} \psi^\ell_{t-k}
%\eeas
%
%{\bf Step 2.3:} generate $\{x^\ell_{t}\}$ as 
%\beas
%\Big\{g_0(x^{\ell, t - p}_{t}, \psi^{\ell, t-q}_{t-1}), \, 1 \le i \le N \Big\}
%\eeas
%
%{\bf Step 2.3:} calculate $\cX^\ell_{1,T}$ from $\{x^\ell_{t}\}$ as in \eqref{eq:test:stat}\;
%}
%{\bf Step 3:} select $\pi_{T}$ as the $100(1-\alpha)$\%-percentile 
%of $\cX^\ell_{1,T} \log(T)^{-1}, \ \ell=1, \ldots, R$\;
%\BlankLine
%\KwOut{$\pi_{T}$}
%\end{algorithm}
However, this resampling procedure adds computational time and it does not fully take advantage of the dampening transformation aiming to bring a series closer to an iid exponential distribution. For that reason we conduct experiments to establish the \emph{universal} value of the threshold parameter under the null hypothesis of no change-points, so that the threshold $\pi_T$ overall depends only on the sample size $T$.  

In particular, we generate stationary ACD processes of size $T$, varying from $500 $ to $100000$ with a step 50. The exact choice of the ACD model parameters (i.e. $(\omega, \alpha, \beta)$) did not alter the results and, therefore, not reported here. Then we find $b$ that maximises (\ref{eq:single_cusum}). The
ratio
$$C_1 = \mathbb{Y}_{s,e}^b  (\sqrt{\log T})^{-1}$$
gives us an insight into the magnitude of parameter $C_1$. We
repeated this experiment 100 times for different values of $T$ and we selected $C_1$ as the $100(1-\alpha)$\%-percentile for each instance of $T$. Our results indicated that $C_1$ tends to decrease as we increase the sample size and remains unchanged after a certain point (see Figure \ref{plot:p_TvsT95and99perc} in the supplementary material).

To propose a general rule that will apply in most cases we fitted the
regression
$$C_1(T)=c_0+c_1 T+ c_2 \frac{1}{T} + c_3 T^2 +\varepsilon.$$
Having estimated the values for $\hat{c}_0, \hat{c}_1,
\hat{c}_2, \hat{c}_3$ we were able to use fitted
values for any sample size $T$. For samples larger than $T=100000$,
we used the same $C_1$ values as for $T=100000$.

We turn to the choice of $M$ - the number of ensembles to run - and $\pi_{thr}$ - the relative frequency threshold. In Section \ref{sec:TheoremEBS} we discussed that the minimum number of $M$ is in the range of a few thousands when the distance of any two adjacent change-points is of $O((\log T)^{1/2})$. However, in practice and thanks to the randomised ensemble mechanism of EBS, $M$ can be much smaller. As for $\pi_{thr}$, the theoretical choice of 0 is adequate, albeit for  the distributional features of the multiplicative setting we consider here a non-zero positive value can work better in practice. Similar to the procedure of the choice of $C_l$ above, we simulate stationary ACD processes and we apply the EBS algorithm for every triplet $(M,\pi_{thr},T)$ keeping everything else the same; see the supplementary material for more details. By inspecting Figure \ref{plot:heatmapsForVaryingSizes}  in the supplementary material, we can see that the choice of $\pi_{thr} = 0.05$ results in a significant improvement in the detection accuracy compared with lower values. For larger values, we do not see any further improvement, even when the sample size increased. In addition we conclude that, conditioning on $\pi_{thr}$, a high number of draws $M$ did not result a considerable improvement in accuracy. On the contrary, a low number in the range of hundreds gave similar results to $M=5000$ even for larger samples. For the simulation study and the real application we set $\pi_{thr} = 0.05$ and $M=500$ which are also the default values in the R package.

\section{Simulation study}\label{sec:Simulat}
We simulated stationary time series with innovations
$\varepsilon_t \sim \exp(1)$ for ACD and sample size $T=2000$ for different specifications. For the Hawkes process we chose the time horizon $h = 500$, whereby the sample size will depend on the choice of the parameters. Roughly speaking, for $\lambda_0=1$, $\alpha_1=0.1$ and $\beta_1 =0.7$ the sample size $T \approx  5000$. We report the false positive rate for each method i.e. the percentage of incorrectly rejecting the null hypothesis of no change-points.

{\small
\begin{table}
\caption{Stationary processes results. For each of the Hawkes processes the average sample size $T_{avg}$ is given in brackets. Figures show
the false positive rate for each method. Average execution time in seconds is given in brackets.}
\label{tab:stationary} \centering {\footnotesize
\begin{tabular}{l|rrrrrr} \hline
\multicolumn{1}{l}{Model}& BS & WBS & EBS  \\
\hline \hline
S1: iid standard poisson with parameters \\ $\lambda_0=2$ ($\alpha_1=\beta_1 = 0 \; \forall j,k$) and $T=2000$ & 2\% (0.018)    & 18\% (0.119) & 4\% (0.061) \\
S2: Hawkes process with parameters \\ $\lambda_0=0.5$, $\alpha_1=0.1$ and $\beta_1=0.7$  ($T_{avg}=250$) & 0\% (0.013)   & 0\% (0.033) & 2\% (0.045)\\
S3: Hawkes process with parameters \\ $\lambda_0=5$, $\alpha_1=0.4$ and $\beta_1=0.7$ ($T_{avg}=5825$) & 22\% (0.337) & 68\% (1.158)  & 26\% (0.415) \\
S4: ACD process with parameters \\ $\omega=1$, $\alpha_1=0.1$ and $\beta_1=0.7$ and $T=2000$ &   4\% (0.017) & 25\% (0.123)  & 8\% (0.066)\\
S5: ACD process with parameters \\ $\omega=3$, $\alpha_1=0.15$ and $\beta_1=0.5$ and $T=2000$  & 9\% (0.180)  & 29\% (0.161) & 9\% (0.063)\\
\hline
\end{tabular}}
\end{table}}

From Table \ref{tab:stationary}, we see that both BS and EBS performed well meaning that the risk of segmenting a stationary process is generally limited. WBS displayed oversegmentation that became more significant such as in cases of processes with high persistence, that is when $\alpha_1+\beta_1$ in the ACD process or the ratio $\alpha_1/\beta_1$ in the Hawkes process is either approaching 1. For example, in the rather challenging case of S3, EBS performed similar to BS, while WBS returned a high false positive ratio (in 68\% of the cases incorrectly rejected the null hypothesis). It is worth mentioning (not shown here) that the risk of oversegmantation increased for WBS when a higher number of random draws was chosen, while the false positive rate for EBS remained unchanged for a larger $M$ (1500 instead of the default 500).
%\subsection{Models with change-points}

We now examine the detection performance of our method for a set of
non-stationary models, both ACD and Hawkes processes. We consider various test models and examine BS and EBS performance over 100 simulations for each of the test model. To compare the performance between the two detection methods we calculate the following error measures: $E(\hat{N}-N)$, $E(|\hat{N}-N|)$, $E[(\hat{N}-N)^2]$. 
%In addition, the EBS method has improved rates of convergence compared with BS, hence, we examine the total number of change-points identified
%within $\lfloor 1\% \cdot T \rfloor$ from the real ones in order to assess the performance on how close the change-points are to their estimates. 

Since EBS has improved rates of convergence compared with BS and it can also assist in the post-processing stage, its accuracy  should be judged in parallel with
the total number of change-points identified. We use a test from \cite{korkas2017}
that tries to accomplish this. Assuming that the maximum
distance from a real change-point $\eta$ is denoted by $d_{\max}$, an
estimated change-point $\hat{\eta}$ is correctly identified if
$|\eta-\hat{\eta}|\leq d_{\max}$ (here within $1\%$ of the sample
size). In the case where two (or more) estimated change-points are within this
distance $d_{\max}$ then only one change-point, the closest to the real
change-point, is classified as correct and the rest are deemed to be
false, except if any of these are close to another change-point.
An estimator performs well when the hit ratio
$$HR = \frac{\# \text{correct change-points identified}}{\max(N,\hat{N})}$$
is close to 1. According to the authors, the term $\max(N,\hat{N})$ 
penalises cases where, for example, the estimator  correctly
identifies a certain number of change-points all within the
distance $d_{\max}$, but $\hat{N} < N$. It also penalises the
estimator when overestimates the total number of change-points and all $\hat{N}$ detected change-points are within distance $d_{\max}$ of the true ones.

We consider nine model specifications M1 to M8.2 ranging from a model with a single change-point to models with 19 change-points (multi-teeth types similar to \cite{fryzwild} for the additive signal + noise setups) and to models with random settings for the number and locations of the change-points. The models are described in detail in the supplementary materials, while the simulation results are provided in Table \ref{tab:nonstationary}. EBS outperformed in many cases both in location accuracy and in controlling the overall number of change-points. What is interesting to note is that the relative performance of EBS against BS or WBS remained high in every scenario and (almost) never falls behind the other two.

{\small
\begin{table}
\caption{\footnotesize{ Nonstationary processes change-point detection results. The error measures are self-explanatory while the hit ratio HR is described in the main text. Average execution time in seconds is given in the last column. In bold, the best or the second best method based on the metric in the respective column is reported.}}
\label{tab:nonstationary}\centering {\footnotesize
\begin{tabular}{ll|rrrrr} \hline
\hline
Method & Model & $E(\hat{N}-N)$ & $E(|\hat{N}-N|)$ &  $E[(\hat{N}-N)^2]$ & HR & exec. time  \\
\hline \hline
&  M1 & \textbf{-0.03} & \textbf{0.03} & \textbf{0.03} & \textbf{0.890} & 0.209 \\
&  M2 & \textbf{-0.11} & \textbf{0.11} & \textbf{0.13} &\textbf{ 0.870} & 0.162\\
&  M3 & 0.97 & 1.13  & 2.35 & 0.674 & 0.014\\
BS&M4 & \textbf{0.67}  & \textbf{1.87}  & \textbf{4.93} & 0.266 & 0.031 \\
&  M5 &\textbf{ 0.21} & \textbf{0.21}  & \textbf{0.31} & \textbf{0.923} & 0.037\\
&  M6 & -6.65 & 7.51  & 76.77& 0.413 & 0.053\\
&  M7 & -10.68 & 10.70 & 135.70& 0.247 & 0.048\\
&  M8.1 &\textbf{-1.14 } &\textbf{1.48} & \textbf{3.70}& \textbf{0.592} & 0.073\\
&  M8.2 &\textbf{-3.82}  &\textbf{ 4.22} & \textbf{24.48}& \textbf{0.535} & 0.083\\
\hline
\hline
&   M1 & -0.18 & 0.18 & 0.32 & 0.775 & 0.621\\
&   M2 & -0.96 & 0.96 & 2.58 & 0.701 & 0.829\\
&   M3 & \textbf{0.09} & \textbf{0.67 }& \textbf{1.13} &\textbf{ 0.719} & 0.112\\
WBS&M4 & 2.27 & 2.51 & 9.33 & 0.327 & 0.371\\
&   M5 & 2.52 & 2.52 & 12.58 & 0.584  & 1.047\\
&   M6 &\textbf{ 5.11} & \textbf{5.17 }& \textbf{33.57} & \textbf{0.593} & 3.710 \\
&   M7 & \textbf{2.90} &\textbf{ 3.42} & \textbf{19.96} & \textbf{0.539} & 1.590\\
&  M8.1 & 1.50 & 2.58 & 13.86& 0.478 & 3.503\\
&  M8.2 &6.67  &8.61 & 119.70& 0.408 & 2.905\\
\hline
\hline
&   M1 & \textbf{0.07} &\textbf{ 0.09} & \textbf{0.09} & \textbf{0.855} & 0.264\\
&   M2 & \textbf{-0.47} & \textbf{0.67} & \textbf{0.47} & \textbf{0.817} & 0.275\\
&   M3 & \textbf{-0.41} & 1.15 & \textbf{2.15} & \textbf{0.686} & 0.069\\
EBS&M4 & \textbf{2.01} & \textbf{2.29} & \textbf{7.77} & \textbf{0.394} & 0.120\\
&   M5 & \textbf{0.38}& \textbf{0.38} & \textbf{0.85} &\textbf{ 0.884}  & 0.484\\
&   M6 &\textbf{ 3.78} & \textbf{4.24} & \textbf{26.00} & \textbf{0.641} & 1.120\\
&   M7 & \textbf{-0.43} & \textbf{2.69} & \textbf{12.59} & \textbf{0.554} & 0.854 \\
&  M8.1 & \textbf{-0.71} & \textbf{1.55} & \textbf{3.91}&\textbf{ 0.552} & 0.791\\
&  M8.2 &\textbf{-2.83}  &\textbf{3.37} & \textbf{16.51}& \textbf{0.617} & 1.017\\
\hline
\end{tabular}}
\end{table}}

\section{Application}\label{sec:Apps}
In this section, we test our proposed methodology using Apple Inc Contract For Differences (CFD) data obtained from Dukascopy for free. CFDs are leveraged products which normally reflect the underlying stock price, but they do not transfer stock ownership to the buyer. They are favoured for their easiness to access (reducing the need to trade in a stock exchange) and the extended trading hours many brokers provide. The data consists of the bid and ask quotes with their timestamps as published by the broker. Alternatively, one could choose transaction data which will require further cleansing due to the existence of multiple small size trades. We analyse all the data from June (20 business days) and from 09:30 until 16:00 (US East local time), the stock exchange trading hours. The daily average sample size (number of quotes per business day) is 60917, the minimum 42007, and the maximum 86777.

For each day in month June we ran EBS (assuming a tvACD process) on the raw durations, the time lapsed between two quotes updates. This is to show whether EBS was able to segment a given series without first removing the intraday seasonality, as a piecewise constant model can approximate it well. Our methodology is designed to capture even small deviations from stationarity. Therefore, it can potentially work well in this setup whereby the underlying dynamics of a duration series change slowly and monotonically at the start of the trading day, flatten during the day and monotonically increase towards the end (resulting in the typical U-shape).

For completeness, we also ran EBS after removing intraday seasonality from the series of durations following \cite{engle1998ACD}. The seasonality effect is defined as a multiplicative component in the following formula
$$\tilde{X}_t = x_t \cdot \phi(\tau)$$
where $\tilde{X}_t$ are the raw durations, $\phi(\tau)$ is the intraday seasonality effect and $x_t$ are the standardized seasonality durations. To estimate $\hat{\phi}(\tau)$ we applied a smoothing spline using cross-validation to select the penalty parameter and the degrees of freedom. On the transformed durations $x_{t}=\tilde{X}_{t} / \hat{\phi}(\tau)$, we applied the EBS algorithm.
 
\begin{figure}[htbp]
\centering
\includegraphics[scale=.5]{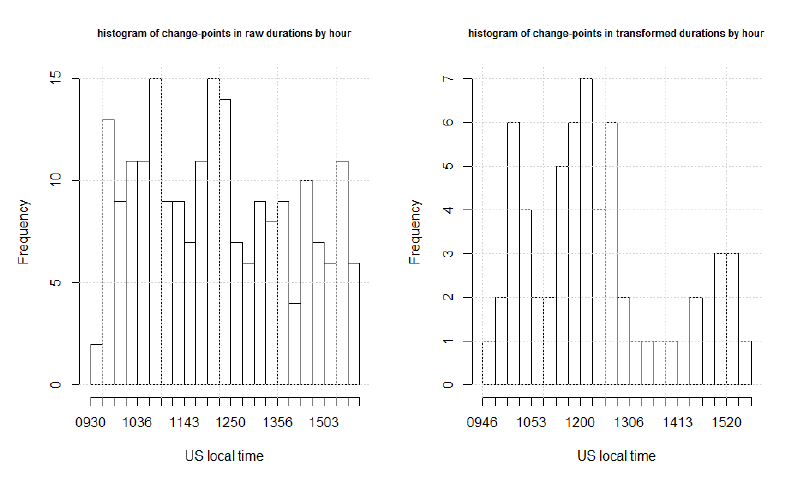}
\caption{\footnotesize{Histogram of detected change-points collected over the sample period grouped by hour. On the left, EBS has been applied on the raw durations and on the right the durations has been transformed such that to remove the intraday seasonality.}}
\label{plot:real_study_hist_comp}
\end{figure}

By inspecting Figure \ref{plot:real_study_hist_comp} below and Figure \ref{plot:Apple_example_durations_vs_adjusted_durations}  in the supplementary material, we see that EBS is able to capture the changes in the trading behaviour during the early and late sessions. Still, when removing seasonality it is very common across the days to observe change-points between 12.00 and 13.00. We do not argue that these change-points capture seasonality left out from applying a smoothing spline. In fact, we believe that financial high-frequency data are characterized by other various types of non-stationarity which cannot be ignored. This observation has implications both for trading and risk management operations. Given the rise of the electronic trading, intraday risk monitoring will need to adjust in order to incorporate previous change-points and provide more reliable forecasts in anticipation of non-stationarity.

\section{Conclusion}
The work has addressed the problem of detecting the change-points
in the structure of an irregularly spaced time series. We proposed a new ensemble-type BS methodology, whereby we combined the change-points detected by applying BS on randomized sub-samples of the data. Doing so, we were able to detect change-points that occur frequently and in short distance between them as well as we controlled the total number of change-points that appeared to be spurious. From the simulation study in Section \ref{sec:Simulat} it appears that the EBS mechanism performs well
at this task. We believe that, apart from the good performance, EBS is a good starting point of studying other ensemble-type change-point detection methods.

%\begin{acknowledgements}
%\end{acknowledgements}


\bibliographystyle{unsrtnat}      % Chicago style, author-year citations
\bibliography{fbib} 
%% Non-BibTeX users please use
%\begin{thebibliography}{}
%
%%
%% and use \bibitem to create references. Consult the Instructions
%% for authors for reference list style.
%\end{thebibliography}

\end{document}